\documentclass[superscriptaddress,nobibnotes,amsmath,amssymb,notitlepage,twocolumn,prl,longbibliography]{revtex4-1}
\usepackage[utf8]{inputenc}

\usepackage{bm,mathptmx,braket}

\usepackage{graphicx,color,hyperref}
\usepackage[caption=false]{subfig}
\hypersetup{colorlinks=true, linkcolor=blue, citecolor=blue, urlcolor=blue} 

\newcommand{\sect}[1]{\vspace{0.3em}{\it #1.}---}

\graphicspath{{./img/}}

\newcommand{\beq}[1]{\begin{equation}\label{#1}}
\newcommand{\eep}{\;.\end{equation}}
\newcommand{\eec}{\;,\end{equation}}
\newcommand{\eeq}{\end{equation}}

\newcommand*\dd{\mathop{}\!\mathrm{d}} 

\newcommand{\la}{\lambda}

\newcommand{\om}{\omega}

\DeclareMathAlphabet{\mathcal}{OMS}{cmsy}{m}{n} 

\renewcommand{\vec}[1]{{\bf #1}}

\newcommand{\kv}{\vec{k}}

\begin{document}

\title{Enhancing the Hyperpolarizability of Crystals with Quantum Geometry}

\date{\today}

\newcommand{\TCM}{{TCM Group, Cavendish Laboratory, University of Cambridge, J.\,J.\,Thomson Avenue, Cambridge CB3 0HE, UK}}
\newcommand{\MCR}{Department of Physics and Astronomy, University of Manchester, Oxford Road, Manchester M13 9PL, UK}

\newcommand{\HarvardSeas}{School of Engineering and Applied Sciences, Harvard University, Cambridge, MA 02138, USA}
\newcommand{\BATH}{Department of Physics, University of Bath, Claverton Down, Bath BA2 7AY, UK}


\author{Wojciech J. Jankowski}
\email{wjj25@cam.ac.uk}
\affiliation{\TCM}

\author{Robert-Jan Slager}
\affiliation{\MCR}
\affiliation{\TCM}

\author{Michele Pizzochero}
\email{mp2834@bath.ac.uk}
\affiliation{\BATH}
\affiliation{\HarvardSeas}

\begin{abstract}
We demonstrate that higher-order electric susceptibilities in crystals can be enhanced and understood through nontrivial topological invariants and quantum geometry, using one-dimensional $\pi$-conjugated chains as representative model systems. First, we show that the crystalline-symmetry-protected topology of these chains imposes a lower bound on their quantum metric and hyperpolarizabilities. Second, we employ numerical simulations to reveal the tunability of nonlinear, quantum geometry-driven optical responses in various one-dimensional crystals in which band topology can be externally controlled. Third, we develop a semiclassical picture to deliver an intuitive understanding of these effects. Our findings offer a firm interpretation of otherwise elusive experimental observations of colossal hyperpolarizabilities and establish guidelines for designing topological materials of any dimensionality with enhanced nonlinear optical properties.
\end{abstract}


\maketitle

\sect{Introduction} Nonlinear optics offers a rich platform to explore a broad range of intriguing phenomena, with implications ranging from fundamental science to engineering~\cite{boyd}. One-dimensional crystals, most notably $\pi$-conjugated linear chains~\cite{Agrawal1978, Dalton1987, Bredas1994}, constitute prime examples to probe enhanced optical responses at nonlinear orders. Because these crystals are centrosymmetric, i.e., they feature an inversion symmetry, their optical responses occur at odd orders in optical fields. The first- and third-order optical responses of one-dimensional $\pi$-conjugated chains have long been investigated, both experimentally ~\cite{Fincher1979, Beratan1987, Sinclair1988, Ifor1994, Ledoux1999, Bredas1994} and computationally~\cite{Hermann1974, Rustagi1974, Hayden1989, Villesuzanne1992, Huang1992, Shuai1992, Shuai1992_2, Schmidt1999, Hu2007, Lacivita2012, Luzanov2023}. Of central interest in this context is the third-order hyperpolarizability, i.e., the third-order electric susceptibility $\chi^{(3)}$ in the static limit of vanishing optical field frequency, $\om \rightarrow 0$, which quantifies the nonlinear polarization response, ${P}^{(3)} = \chi^{(3)} E_x^3$, to electric field $E_x$~\cite{Pronin1995}. In $\pi$-conjugated chains, $\chi^{(3)}$ is known to acquire colossal values. For example, $\chi^{(3)}$  was inferred to be as high as  $10^{-18}$$~$m$^2$/V$^2$ in polyacetylene~\cite{boyd},  $10^{-19}$$~$m$^2$/V$^2$ in polydiacetylene, and $10^{-20}$$~$m$^2$/V$^2$ in polyarylenes~\cite{Bredas1994}. This unusually large hyperpolarizability has been heuristically ascribed to the responses of $p_z$ electrons forming the conjugated $\pi$ bonds~\cite{boyd}. However, a general theoretical understanding of the phenomenological origin of these nonlinear optical properties remains yet to be achieved~\cite{Yu1989_2, Hagler1994, Takahashi1995, Yaron1995, Takahashi1997, Takahashi1977_2, Matsuzaki1998, Margulis1998, Jiang2005, Xu2005, DiazPonce2017}.

\begin{figure}[t!]
\includegraphics[width=0.9\columnwidth]{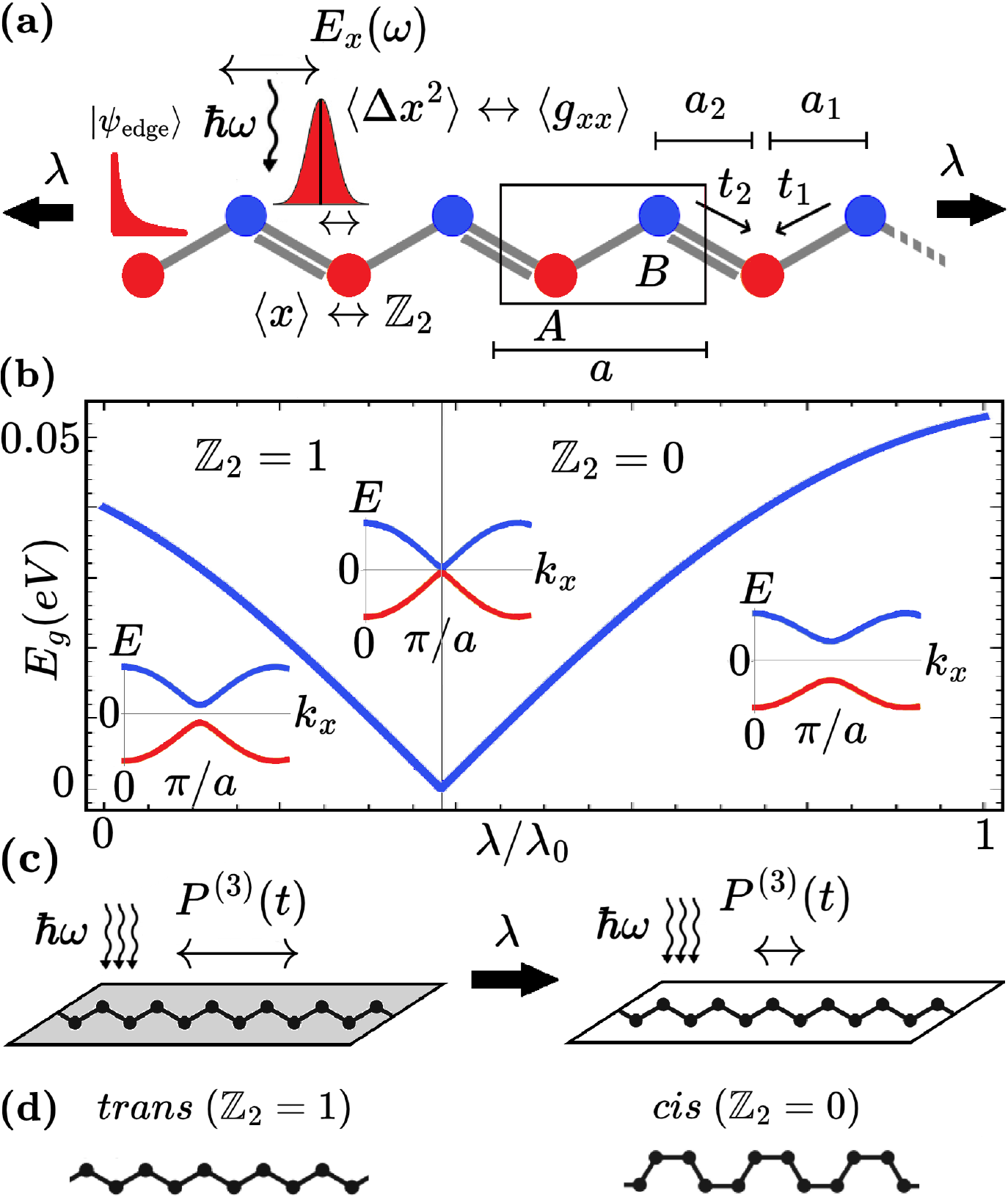}
\caption{
{Crystalline topology and quantum geometrically enhanced hyperpolarizability in one-dimensional crystals.} { {(a)} } Schematic illustration of a one-dimensional inversion-symmetric model system with a topological invariant $\mathbb{Z}_2$ reflected by a Wannier charge center shift $\langle x \rangle$ and by the presence of a symmetry-protected exponentially localized edge state, $\ket{\psi_{\text{edge}}}$. The model system with unit cell parameter $a$ realizes a bipartite basis $A,B$, and effective hoppings $t_1, t_2$ across bonds with lengths $a_1, a_2$, and is coupled to an optical field $E_x(\om)$ introduced by photons with energy $\hbar \om$. The minimal spread of the Wannier function $\langle \Delta x^2\rangle$ is topologically bounded, and evaluates to the band-averaged quantum metric $\langle g_{xx} \rangle$~\cite{Marzari1997}, which determines the nonlinear polarizability. { {(b)} } Control of the band gap and band topology with lattice strain  ($\la$), across a topological phase transition (TPT), where $\lambda_0$ is a unit strain. { {(c)} } Discontinuous control of the nonlinear optical properties, such as the time-dependent third-order polarization $P^{(3)}(t)$ from a three-photon coupling, across a~strain-induced TPT. {\bf {(d)}} Molecular structure of the \textit{trans} and \textit{cis} isomers of polyacetylene realizing different $\mathbb{Z}_2$ invariants, resulting in distinct hyperpolarizabilities.}
\label{Fig1}
\end{figure}

\begin{figure*}[t!]
\includegraphics[width=\linewidth]{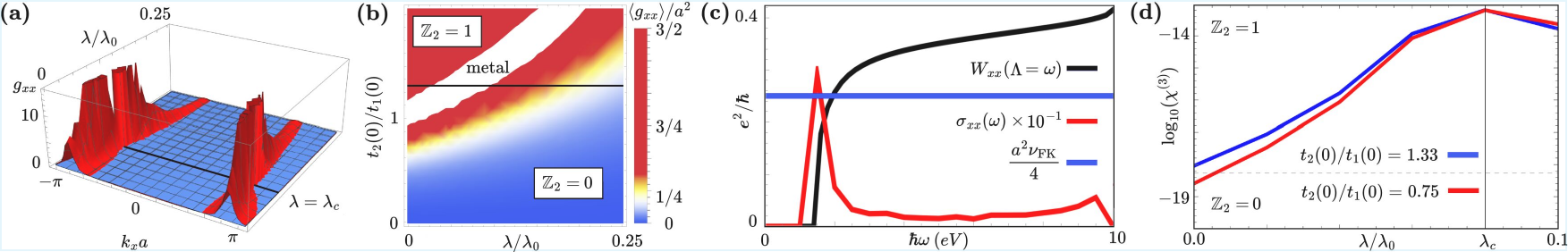}
\caption{{Quantum geometry and (non)linear optics in one-dimensional crystals.} { (a) } Quantum metric ({red}) over momentum space as a function of the lattice strain parameter $(\lambda)$ realized in the effective low-energy Hamiltonian, Eq.~\eqref{eq:strain}, with hoppings $t_1(0) = 2.15~\text{eV}$, $t_2(0) = 2.85~\text{eV}$ [$t_2(0)/t_1(0) = 1.33$]. The critical strain $(\lambda = \lambda_c)$ induces a topological phase transition across which the metric scales up at the band edge at $k= \pm\pi/a(\la)$, overcoming the minimal average metric value per BZ ({blue}) necessary for satisfying the topological bound. { {(b)} } Band-averaged metric $\langle g_{xx} \rangle$ as a function of strain and hopping amplitude ratio $t_2(0)/t_1(0)$. The horizontal line corresponds to the chain with $t_2(0)/t_1(0) = 1.33$. The corresponding optical absorptivity is manifestly larger and topologically lower bounded in the topologically nontrivial phase.  { {(c)} } Spectral resolution of strain-dependent optical conductivity $\sigma_{xx}(\om)$ and cutoff ($\Lambda$) dependent optical weights $W_{xx}(\Lambda) = \int^\Lambda_0~\dd\om~ \frac{\sigma_{xx}(\om)}{\om}$ ({black}), as a function of the bounding value for $\Lambda \rightarrow \infty$ ({blue}) in a topological chain with $t_2/t_1 = 1.33$.  { {(d)} } Lattice strain control of the topologically bounded hyperpolarizability $\chi^{(3)}$.   In the subcritical strain limit, we find that due to distinct realized quantum metrics, the $\chi^{(3)}$ of a trivial phase [$\mathbb{Z}_2 = 0$] ({red}) is multiple times smaller than the $\chi^{(3)}$ of a topological phase $\mathbb{Z}_2 = 1$] ({blue}) with identical band energies and band gaps, which determine equal $\lambda_c$ for transitions to metallic states. The dashed line denotes the topological lower bound on $\chi^{(3)}$. Under the relatively high values of strain $\la/\la_0$, where $\lambda_0$ is a unit strain, the anomalous hyperpolarizability of the $\mathbb{Z}_2$-topological phase can be quenched; see also the Supplemental Material~\cite{SI}. At vanishing strain, we note that the effective model fitted to the $t_2/t_1$ of polyacetylene yields \mbox{$\chi^{(3)} = 8 \times 10^{-19} $m$^2$/V$^2$}, in excellent agreement with the corresponding experimental value, $\chi^{(3)} \approx$ 10$^{-18}$~m$^2$/V$^2$~\cite{Fincher1979, Dalton1987, Bredas1994}.}
\label{Fig2}
\end{figure*}

In this Letter, we bridge this gap by deploying the modern notions of band topology and quantum geometry. We show that the crystalline-symmetry-protected topology of $\pi$-conjugated chains induces a lower bound on the quantum metric, and thus on the hyperpolarizability. Then, we use numerical simulations to demonstrate  the~tunability of nonlinear, quantum geometry-driven optical responses of a variety of one-dimensional crystals in which the band topology is controllable by an external parameter, such as lattice strain. Finally, we propose a general semiclassical picture to explain these remarkable optical properties, based on the scaling of quantum metric and band gaps with external parameters. Importantly, our theory offers a topological explanation for the distinct hyperpolarizabilities of the \textit{cis} and \textit{trans} isomers of $\pi$-conjugated linear polymers shown in Fig.~\ref{Fig1}(d)~\cite{Sinclair1988}. Unlike the \textit{cis} configuration, the topological windings of the Bloch eigenvector in the \textit{trans}  configuration support the geometric enhancements of the optical matrix elements associated with the nonlinear optical responses. Overall, our findings provide a novel interpretation of otherwise elusive experimental observations, along with guiding principles for identifying and engineering materials for nonlinear optics. 

\sect{Quantum geometry and topology in one dimension} Over the past several years, crystalline topologies have been thoroughly investigated~\cite{fu2011,rjs_translational, Slager2013, kruthoff2017, Clas4, Clas5}, culminating in a rather uniform view upon comparing momentum space constraints~\cite{kruthoff2017} and real space (Wannierizability) conditions~\cite{Clas4,Clas5}. Within this framework, taking into account the fact that an exponentially localized Wannier basis~\cite{Marzari2012} can be always constructed~\cite{Kohn1959} in one spatial dimension, it is directly inferred that the only form of nontrivial electronic topology  is that of an obstructed limit. The topology~\cite{Schnyder2008, Kitaev2009} arises from the interplay of band wave function ($\ket{\psi_{n\kv}} = e^{i\kv\cdot \vec{r}} \ket{u_{n\kv}}$) windings with additional symmetries, such as chiral or inversion ($\mathcal{I}$) symmetry~\cite{Fu2007}. Under $\mathcal{I}$~symmetry, in nonmagnetic systems, the topological invariant is determined by the band parity eigenvalues, $\delta_k = \pm1$ at inversion-symmetric momenta, $k = 0, \pi/a$, where $a$ is the lattice constant, which coincides with the Fu-Kane formula~\cite{Fu2007}: ${\nu_{\text{FK}} = (1-\delta_0 \delta_{\pi/a})/2 \in \mathbb{Z}_2}$. At the same time, the band topology of obstructed insulators provides for the minimal spread $\langle \Delta x^2 \rangle$ of the exponentially localized Wannier functions~\cite{Marzari1997}, as schematically presented in Fig.~\ref{Fig1}(a). Furthermore, this can be understood with the notions of quantum geometry~\cite{provost1980riemannian}, and can be related to topological bounds~\cite{Jonah2022}.

We now elucidate how, in one-dimensional (1D) crystals, the crystalline topology gives rise to quantum geometry and, as a direct consequence, enhances their optical responses. We define a non-Abelian multiband quantum geometric tensor, ${[\textbf{Q}^{(m)}_{ij}]_{np} = \braket{\partial_{k_i} u_{n\kv}|u_{m\kv}} \braket{u_{m\kv}|\partial_{k_j} u_{p\kv}}}$ over momentum space, with its real part, commonly known as the non-Abelian quantum metric, $[\textbf{g}^{(m)}_{ij}]_{np} = \text{Re}~[\textbf{Q}^{(m)}_{ij}]_{np}$~\cite{provost1980riemannian, Ma2010, bouhon2023quantum}. The multiband quantum metric encodes the optical transition dipole matrix elements, $[\textbf{g}^{(m)}_{xx}]_{nn} = |\bra{\psi_{n\kv}} \hat{x} \ket{\psi_{m\kv}}|^2$~\cite{Ahn2020, Ahn2021}, which is at the heart of both linear and nonlinear responses. The multiband quantum metric of electrons realizes lower bounds due to 1D crystalline topologies that translate into minimal values of (higher-order) optical susceptibilities,
\beq{eq:bound}
   \langle g_{xx} \rangle \equiv \sum^{\text{unocc}}_{m} \sum_{k} \text{Tr}~{\textbf{g}}^{(m)}_{xx} \geq \frac{a^2}{4} \nu_{\text{FK}},
\eeq
where $\langle g_{xx} \rangle$ is band-averaged quantum metric and ${\nu_{\text{FK}} \in \mathbb{Z}_2}$ the Fu-Kane parity invariant protected by the {$\mathcal{I}$~symmetry}~\cite{Fu2007}. The bound can be derived from (i) nontrivial inversion symmetry indicators pinpointing how the irreducible representations of the obstructed bands transform~\cite{Fu2007, kruthoff2017, Clas4, Clas5}, (ii) translating these into real space projectors~\cite{Jonah2022}, and (iii) using an optimization argument of Ref.~\cite{Jonah2022}. The technical details of the derivation are given in Sec.\ I of the Supplemental Material~\cite{SI}. In the context of the maximally localized Wannier functions, $\langle g_{xx} \rangle = \langle{\Delta x^2} \rangle$~\cite{Marzari1997}. Hence,  the invariant imposes a lower-bound constraint on the spread of the wave function in the real space basis. In topological crystalline phases, any topologically nontrivial band realizes $\nu_{\text{FK}} = 1$. In the following, we demonstrate that this bound yields a topological enhancement of the multiband geometry-dependent response functions within the first- and third-order optical responses.

\sect{First- and third-order optical responses} The linear optical conductivity of a band insulator in the first-order response to an electric field of frequency $\om$ reads as~\cite{Ahn2021},
\beq{}
    \sigma_{xx} (\om) = \frac{e^{2}}{2\hbar^2} \int_\text{BZ} \frac{\dd k}{2\pi} \sum_{n,m}~f_{nm} E_{mn} \text{Tr}~{\textbf{g}}^{(m)}_{xx} \delta(\om - E_{mn}/\hbar),
\eeq
where $f_{nm} = f_{nk} - f_{mk}$ are the thermal Fermi-Dirac occupation factors, momentum $k$ runs over the first Brillouin zone (BZ), and $E_{mn}$ is the difference in energy between valence and conduction bands $n$ and $m$, respectively. For crystals possessing the same band structure, the distinct first-order optical conductivity is purely determined by the multiband quantum metric. This allows one to uniquely distinguish band topologies by inducing the corresponding band geometries. Importantly, changes in band topology, as realized through external parameters such as lattice strain, as shown in Figs.~\ref{Fig1}(a) and~\ref{Fig1}(b), result in changes in the optical response. As we shall discuss in the representative case of $\pi$-conjugated chains and related one-dimensional crystals, lattice strain can be leveraged to engineer the opacity and optical nonlinearities, as illustrated in Fig.~\ref{Fig1}(c).

While the second-order optical responses vanish under the $\mathcal{I}$~symmetry in centrosymmetric crystals, third-order responses remain finite and can be controlled  with band geometry. 
In the static limit, the third-order susceptibility can be enhanced quadratically with the considered multiband quantum metric
\beq{eq:chi3}
    \chi^{(3)} = \frac{e^4}{\epsilon_0} \int_\text{BZ} \frac{\dd k}{2\pi}~ \sum_{n,m,p,q} \frac{[\textbf{g}^{(p)}_{xx}]_{nm} [\textbf{g}^{(q)}_{xx}]_{mn} }{E_{mn}E_{pn}E_{qn}},
\eeq
where $\epsilon_0$ is the vacuum permittivity.  The derivation of Eq.~\eqref{eq:chi3} is provided in Sec.\ II of the Supplemental Material~\cite{SI}.

While the enhancement of $\chi^{(3)}$ can be induced by nontrivial quantum geometry in the numerator, it can be further achieved through the tunable band gap magnitudes in the denominator. Notably, materials with identical band gaps, yet with distinct crystalline topologies, realize distinguishable numerators, with topological phase experiencing a quadratic enhancement. By combining this enhancement with Eq.~\eqref{eq:bound} and sum rules, we derive a fundamental topological lower bound on the hyperpolarizability of a nontrivial insulator, 
\beq{eq:lb}
\chi^{(3)} \geq \alpha^{(3)} \times \frac{a^{10}}{n_e^3} ,
\eeq
where $\alpha^{(3)} = \frac{1}{2}\pi^3 e^4 m_e^3 \epsilon^{-1}_0 \hbar^{-6} = 6 \times 10^{50}~\text{m}^{-5}/\text{V}^2$, $m_e$ is the electron mass, and $n_e$ is the number density of electrons in the crystal. The derivation of Eq.~\eqref{eq:lb} employs (i) the energy gap bound of Ref.~\cite{Kivelson1982} due to quantum metric (ii)
combined with the sum rules of Ref.~\cite{Souza2000}, as employed in the linear susceptibility bound of Ref.~\cite{Onishi2024}. The technical details of the derivation are given in Sec. IV of the Supplemental Material~\cite{SI}.

\sect{From models to real materials}
We next employ an effective, two-band Hamiltonian to capture the geometry induced by the $\mathbb{Z}_2$ obstructed topology in the presence of lattice strain. This model is schematically depicted in Fig.~\ref{Fig1}(a). The Hamiltonian that minimally realizes such general phenomenology on a quasi-one-dimensional chain takes the form of a strained Su-Schrieffer-Heeger (SSH) model~\cite{ssh1, ssh2},
\\
\begin{small}
\beq{eq:strain}
    H(k_x, \lambda) = \begin{pmatrix}
    0 & t_1(\lambda)  + t_2(\lambda)  e^{-i k_x a(\la) }\\
    t_1(\lambda)  + t_2(\lambda)  e^{i k_x a(\la)}  & 0 \\
    \end{pmatrix},
\eeq
\end{small}
\\
where $\lambda > 0$ is an external tensile stress parameter, ${a(\la) \equiv a_1(\la) + a_2(\la)}$ is the strain-renormalized lattice parameter with bond lengths $a_{j}(\la) = a_{j}(0)[1+ \la]$, and ${t_{j}(\la) = t_{j}(0)e^{-\gamma_j[a_{j}(\la) - a_{j}(0)]}}$ are the strain-dependent hopping amplitudes, where $\gamma_j$'s parametrize bond stiffness~\cite{Tepliakov2023}. We validate our strain-dependent models with hopping parameters obtained from first-principles calculations \cite{jankowski2024excitonic}. 
We correspondingly present our numerical results in Fig.~\ref{Fig2}. In Fig.~\ref{Fig2}(a), we show how the profile of quantum metric $g_{xx}$ over $k$~space changes with the strain $\la$. The metric supporting the topological lower bound is enhanced at the band edge $k = \pi/a$, as the dominant contributions arise at the band inversion region, where the local derivatives of the Bloch eigenvectors winding across the BZ are maximized. Upon crossing the critical strain $\la = \la_c$, the band edge metric contributions provide for an average $\langle g_{xx} \rangle$ that satisfies the bound in the topological phase ($\nu_{\text{FK}}=1$). Figure~\ref{Fig2}(b) presents a dependence of the optical weight, i.e., weighted integrated optical conductivity, or equivalently, integrated interband contribution to the imaginary part of the dielectric constant ${W_{xx} =\int^\infty_0~\dd \om~ \frac{\sigma_{xx}(\om)}{\om} = (\pi e^2 / \hbar a) \langle g_{xx} \rangle}$, which is also lower bounded by the topological invariant. In.~Fig.~\ref{Fig2}(c), we benchmark the first-order optical conductivity and show how its contributions to the bounded optical weights gradually arise. Figure~\ref{Fig2}(d) shows how the colossal values of $\chi^{(3)}$ arise with strain $\la$, before vanishing sharply upon transitioning to the trivial phase across the critical point. We remark that the value of $\chi^{(3)}$ at vanishing $\la$ is in excellent agreement with experimental observations. Importantly, the bound of the quantum metric, as manifested in Figs.~\ref{Fig2}(c) and~\ref{Fig2}(d), provides for the robustness of lower bounded optical responses of topological phases, especially away from the singular critical points that correspond to the strain-induced phase transitions. Further details on the model-dependent parametrization and controllability of quantum geometry within \mbox{$\pi$-conjugated} chains are given in Sec.~III of the Supplemental Material~\cite{SI}.

\begin{figure}[t!]
\includegraphics[width=\columnwidth]{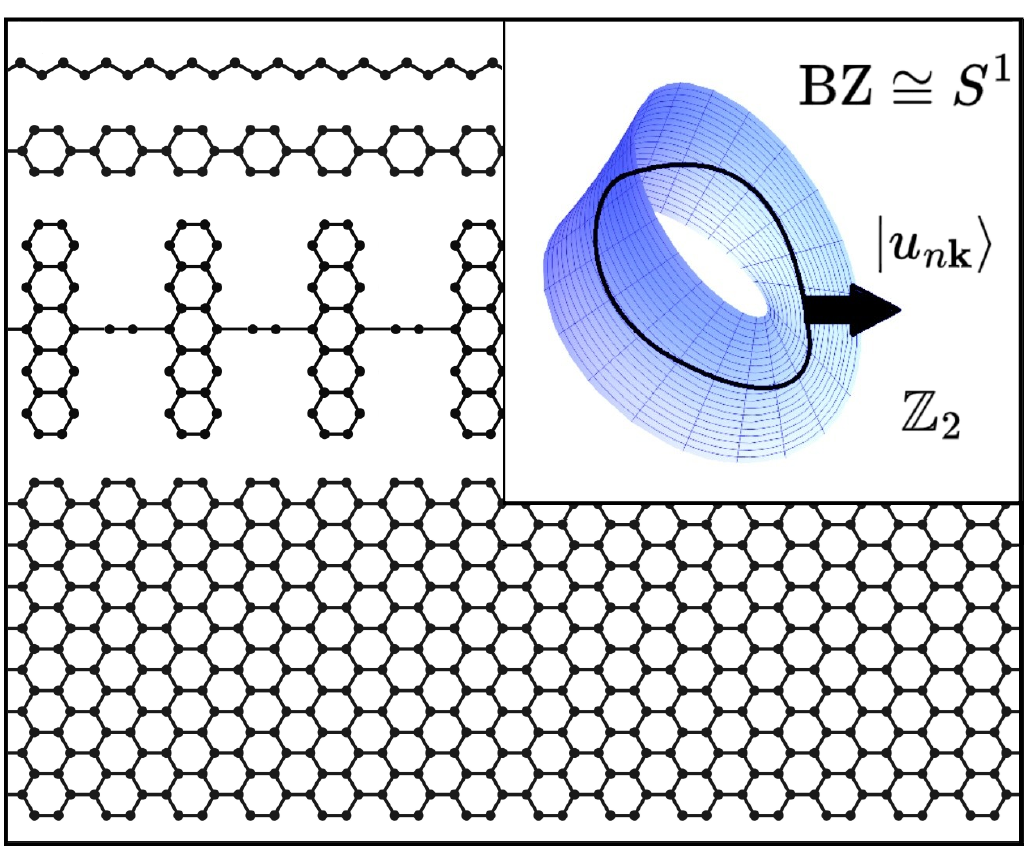}
\caption{
{Examples of materials for crystalline topology-induced, quantum-geometrically enhanced nonlinear optics.} From top to bottom: polyacetylene, poly-($p$-phenylene), polypentacene, and graphene nanoribbons. Under external control, such as lattice strains, these materials can realize $\mathcal{I}$~symmetry protected nontrivial $\mathbb{Z}_2$ topology and topologically enhanced hyperpolarizability $\chi^{(3)}$, which culminates in nonlinear polarization $P^{(3)}$. Further chemical functionalizations of such $\mathbb{Z}_2$-topological materials, e.g., through perhalogenations, can provide a further knob to tune the quantum metric and higher-order susceptibilities.}
\label{Fig3}
\end{figure}

We stress that this model Hamiltonian and accompanying results are effectively realized by the low-energy bands of an ample variety of one-dimensional $\pi$-conjugated systems, such as polyacetylene~\cite{ssh1, ssh2}, polyacenes~\cite{Cirera2020}, polyphenylenes, and graphene nanoribbons~\cite{Tepliakov2023}, as overviewed in Fig.~\ref{Fig3}. Their low-energy valence and conduction bands are universally governed by an effective physics of hopping between the Wannier orbitals $A,B$ that are predominantly constituted by the weighted combinations of the atomic orbitals with $p_z$ character~\cite{ssh1, Cirera2020, jankowski2024excitonic}. The physical implication of our findings is that the materials shown in Fig.~\ref{Fig3} are expected to exhibit quantum-geometrically enhanced hyperpolarizabilities that originate from their nontrivial, inversion symmetry-protected band topology. Indeed, the introduction of additional bands does not alter our conclusions, as we demonstrate in Sec.~V of the Supplemental Material~\cite{SI}, where we also provide an extended discussion of multiband models. The reason for this traces back to the fact that the additional bands will experience higher energy gaps $E_{mn}$, suppressing the $\chi^{(3)}$ contributions as $1/ E^3_{mn}$. Furthermore, the metric similarly scales with $1/E^2_{mn}$ [cf.~Fig.~\ref{Fig2}(a)], which for topologically trivial additional bands provides only a marginal contribution to the numerator of $\chi^{(3)}$.

\sect{From quantum to semiclassical picture} We now present an intuitive picture of the topologically induced quantum geometric phenomenology. Semiclassically, electrons in band gap insulators can be modeled with an effective nonlinear anharmonic oscillator Hamiltonian, ${H = \frac{p^2}{2m_e} + U(x)}$ with a $\mathcal{I}$-symmetric confining nonlinear potential ${U(x) = \frac{1}{2} \kappa x^2 + \frac{1}{4} b x^4}$, where $\kappa$ is an effective spring constant, $\kappa = m_e\om_0^2$, in which $\om_0$ is a resonant frequency and $b$ quantifies the degree of anharmonicity. Notably, ${\kappa ~\propto 1/\langle g_{xx} \rangle = 1/\langle \Delta x^2 \rangle}$~\cite{Onishi2024}, where $\langle \Delta x^2 \rangle$ is a spatial spread of a localized state wave function, as representable in terms of Wannier functions~\cite{Marzari1997}. From the theory of the nonlinear anharmonic oscillator discussed in Sec.\ VI of the Supplemental Material~\cite{SI}, one expects $\chi^{(3)} \propto b/\kappa^4$, by recognizing that, typically, $b \propto \kappa/ \langle g_{xx} \rangle$, explaining the effective scaling of the ${\chi^{(3)} \propto \langle g_{xx} \rangle^2}$ within a full quantum-mechanical treatment. The semiclassical picture, in connection to quantum geometry, further supports the interpretation of our key result. Namely, the topological symmetry-indicated winding of the Bloch states in a one-dimensional insulator (i) induces a geometric spread in the localization of the bounded electrons, which (ii) semiclassically reflects an effective spring constant and its inteplay with anharmonicity. The latter determine the nonlinear optical response of the driven localized electrons within an anharmonic oscillator picture.

\sect{Discussion and conclusions} In summary, we have shown that the colossal linear and higher-order optical responses observed in one-dimensional  $\pi$-conjugated chains can be understood by means of the modern notions of band topology and quantum geometry, where the latter is bounded from below by the former. We have established a transparent connection of the multiband quantum geometry with the third-order optical responses, demonstrating that the crystalline-symmetry-protected topology induces a lower bound on the hyperpolarizability. Our quantum-geometric perspective sheds a new light on the long-standing controversy of the discrepancy between $\chi^{(3)}$ of the $\textit{cis}$ and $\textit{trans}$ geometric isomers of polyacetylene~\cite{Sinclair1988, Yu1989, Bredas1994}. The distinction between \textit{cis}- and \textit{trans}-polyacetylene arises from the different spatial arrangements of atoms with identical chemical connectivity, which induces inequivalent band topology in the two isomers. The band topology
classified by the winding of the Bloch eigenvectors provides for a lower bound in the quantum geometric tensor
and in the associated enhanced nonlinear optical response matrix elements in the \textit{trans} isomer. While being manifestly larger in the topological polymer, the metric band
average lies above the topological lower bound and does not saturate, unless the band dispersion is forced
to approach the flat band limit. The minimization of the lower bounded quantum metric in such a limiting case
is consistent with the other known flatness conditions for the metric bound saturations~\cite{Roy2014, Kwon2024, Wahl2025}. In the provided picture, $\textit{cis}$-polyacetylene is topologically trivial (${\nu_{\text{FK}} = 0}$), hence, realizes no bound on geometry with the average quantum metric $\langle g_{xx}\rangle$ that induces an enhancement of the $\chi^{(3)}$, unlike in the topological $\textit{trans}$-polyacetylene ($\nu_{\text{FK}} = 1$). 

The theory presented in this Letter offers a step forward by revealing that the origin of the $p_z$-orbital band polarizability is topologically induced, as associated with the lower bounded quantum metric providing for an enhanced spread of the electronic wave function in the topological phase. To the best of our knowledge, the importance of band topology and geometry for nonlinear responses in the static limits--- especially concerning hyperpolarizabilities----was not addressed in previous studies. Given that any band topology is fully encoded in the non-Abelian Berry connections, which are fully captured by the quantum geometry determining nonlinear optical responses~\cite{Ahn2021}, our conceptual framework consistently and comprehensively captures the role of band topology for hyperpolarizabilities in any class of materials. In addition to offering an understanding of the observed difference in the hyperpolarizabilities between distinct isomers through the notion of quantum geometric bound, the phenomenological picture underlying our findings translates into topologically induced quantum lattice fluctuations \cite{Yu1989_2, Takahashi1995,Takahashi1977_2} in the form of virtual soliton-antisoliton pairs~\cite{Yu1989_2}. Following identifications of positional charge correlators with electronic localization lengths and quantum geometry of Refs.~\cite{Kivelson1982, Souza2000, PokMan2024}, such quantum charge density fluctuations are driven by a quantum metric.

To conclude, our Letter demonstrates that topologically lower bounded higher-order geometric optical responses can be achieved in materials possessing nontrivial crystalline topology, thus opening a pathway to identify and engineer novel platforms for enhanced nonlinear phenomena within the growing family of topologically classified materials.\\

\sect{Acknowledgements} W.J.J. and \mbox{R.-J.S.} acknowledge Bartomeu Monserrat and Joshua J.P. Thompson for stimulating discussions. W.J.J.~acknowledges funding from the Rod Smallwood Studentship at Trinity College, Cambridge. R.-J.S. acknowledges funding from an EPSRC ERC underwrite Grant No.~EP/X025829/1 and a Royal Society exchange Grant No.~IES/R1/221060 as well as Trinity College, Cambridge.

\bibliography{references}

\end{document}


\title{{\textsc{supplemental material}} \\ Enhancing the Hyperpolarizability of Crystals with Quantum Geometry}

\newcommand{\TCM}{{TCM Group, Cavendish Laboratory, University of Cambridge, J.\,J.\,Thomson Avenue, Cambridge CB3 0HE, UK}}
\newcommand{\MCR}{Department of Physics and Astronomy, University of Manchester, Oxford Road, Manchester M13 9PL, UK}

\newcommand{\HarvardSeas}{School of Engineering and Applied Sciences, Harvard University, Cambridge, MA 02138, USA}
\newcommand{\BATH}{Department of Physics, University of Bath, Claverton Down, Bath BA2 7AY, UK}


\author{Wojciech J. Jankowski}
\email{wjj25@cam.ac.uk}
\affiliation{\TCM}

\author{Robert-Jan Slager}
\affiliation{\MCR}
\affiliation{\TCM}

\author{Michele Pizzochero}
\email{mp2834@bath.ac.uk}
\affiliation{\BATH}
\affiliation{\HarvardSeas}

\date{\today}

\maketitle

\begin{widetext}

\tableofcontents 

\newpage
\section{Multiband geometry and topological bounds}\label{app::A}

We provide further details on the multiband geometry and topological bounds central to the main text. The band topology is geometrically encoded in the non-Abelian Berry connection $\vec{A}_{nm} = i\bra{u_{n\kv}}\ket{\nabla_\kv u_{m\kv}}$, with the energy eigenstates of the Bloch wave form $\ket{\psi_{n\kv}} = e^{i\kv \cdot \vec{r}} \ket{u_{n\kv}}$. Furthermore, the non-Abelian Berry connection determines the optical responses of materials, with transition dipole matrix elements $\vec{r}_{nm} = (1-\delta_{mn}) \bra{\psi_{n\kv}} \vec{\hat{r}} \ket{\psi_{m\kv}} = i(1-\delta_{mn}) \vec{A}_{nm}$ between occupied bands $n$ and unoccupied bands $m$~\cite{Ahn2021}.  The non-Abelian Berry connection determines the multiband metric~\cite{Ahn2020},
\beq{}
    g^{nm}_{ij} = A^i_{nm} A^{j}_{mn}   = \text{Re}~\bra{\partial_{k_i} u_{n\kv}}\ket{u_{m \kv}} \bra{u_{m \kv}}\ket{\partial_{k_j}  u_{n\kv}}, 
\eeq
which is a gauge-independent quantity, given it can be rewritten as $g^{nm}_{ij} = \frac{1}{2} \text{Tr}~[(\partial_{k_i} P_n)(\partial_{k_j} P_m)]]$~\cite{Ahn2021}, with $P_l = \ket{u_{l\kv}} \bra{u_{l\kv}}$ gauge-independent projector onto a band $l$. Hence, as a gauge-independent quantity, the multiband quantum metric can indeed be connected to the physical response functions, and in particular to optical signatures. In the context of the most general non-Abelian multiband quantum metric, here we specifically consider $g^{nm}_{xx} = [\mathbf{g}_{xx}^{(m)}]_{nn}$, with $[\mathbf{g}_{xx}^{(m)}]_{np} = A^x_{nm} A^{x}_{mp}$, which is consistent with the main text definition in terms of momentum space tangent vectors $\partial_{k}$ and associated geometry induced by the dreibeins $\{ \ket{u_{n\kv}},\ket{u_{m\kv}}, \ket{u_{p\kv}} \}$.

On summing over all unoccupied bands, as in the main text, one can retrieve a topological bound, as we show below:
\beq{}
    g^{n}_{xx} \equiv \sum^{\text{unocc}}_{m} g^{nm}_{xx} \geq \frac{a^2}{4} \nu^n_{\text{FK}},
\eeq
with $\nu^n_{\text{FK}} \in \mathbb{Z}_2$, the crystalline Fu-Kane (FK) invariant~\cite{Fu2007} for inversion-symmetry protected topology of an isolated band $n$ in one spatial dimension. In that context, in terms of the non-Abelian Berry connection, the FK invariant can be written as~\cite{Hughes2011},
\beq{}
    \nu^n_{\text{FK}} = \frac{1}{2\pi} \int_{\text{hBZ}} \dd k~ \Big[ A^x_{nn}(k) + A^x_{nn}(-k) \Big],
\eeq
with $\text{hBZ}$ the irreducible half of a conventional Brillouin zone. Below, we provide the general derivation of the topological bound in the one-dimensional inversion-symmetric systems. The argument is analogous to the cases of obstructed insulators in higher dimensions~\cite{Jonah2022}. 
To proceed, we begin by noting that the $k$-space irreps of the occupied bands are constituted by the $\mathcal{I}$-symmetry eigenvalues at BZ points $\Gamma$ ($k=0$) and $X$ ($k=\pi/a$). the multiplicities $m$ of irreps satisfy:
\beq{eq:multi}
    m(X) - m(\Gamma) = \text{Tr}~D(\mathcal{I}) P(\Gamma),
\eeq
with a representation matrix $D(\mathcal{I}) = \text{diag}(1,-1)$ of $\mathcal{I}$-symmetry acting on the orbitals, and $P(k)=\sum^{\text{occ}}_n P_n(k)$ a projector onto the occupied bands. Furthermore,
\beq{}
    P(X) - P(\Gamma) =  2 \sum^2_{n=1} p(-\mathcal{I}^n a) + (x > a)~\text{terms},
\eeq
with $p(x) \equiv \int_\text{BZ} \frac{dk}{2\pi} e^{ikx} P(k)$, which follows definitionally. On taking a Frobenius norm and triangle inequality~\cite{Jonah2022}, we moreover have:
\beq{}
    ||P(X) - P(\Gamma)|| \leq ||p(-a)|| + ||p(a)|| + ||(x > a )~\text{terms}||,
\eeq
and furthermore $||p(a)|| = ||p(-a)||$, directly from definition. From the same definition,
\beq{}
    \langle g_{xx} \rangle = \frac{a}{2\pi} \int_\text{BZ}~dk~\frac{1}{2}~\text{Tr}[(\partial_{k} P)(\partial_{k} P)] = \sum_{x \in ja} |x|^2 ||p(x)||^2 \geq a^2 ||p(a)||^2,
\eeq
with $j \in \mathbb{Z}$. We employ the inequality $||P(X) - P(\Gamma)||^2 \geq \text{Tr}~||D(\mathcal{I})(P(X) - P(\Gamma))||/2$, which here holds specifically for matrices with rank $2$, and also holds for any other unitary matrix than $D(\mathcal{I})$, similarly to the conditions used in Ref.~\cite{Jonah2022}. As a consequence, by an analogous optimization argument to Ref.~\cite{Jonah2022},
\beq{}
    \sum_{x \in ja} |x|^2 ||p(x)||^2 \geq a^2 \text{Tr}~||D(\mathcal{I})(P(X) - P(\Gamma))|| = \frac{a^2}{4} \nu_{\text{FK}},
\eeq
with the last relation following directly on inserting the multiplicity relations Eq.~\eqref{eq:multi}. This relation concludes the derivation of the bound,
\beq{}
    \langle g_{xx} \rangle \geq \frac{a^2}{4} \nu_{\text{FK}},
\eeq
which for isolated bands can be generalized to individual single-band bounds, on sending: $g_{xx} \rightarrow  g^n_{xx}$, $P \rightarrow  P_n$, $p(x) \rightarrow  p_n(x)$, $\nu_{\text{FK}} \rightarrow \nu^n_{\text{FK}}$, and on repeating exactly the same derivation (as above) under such substitution.

Importantly, the nontrivial FK invariant, coinciding with the parity index, $\nu_{\text{FK}} = 1$, can be realized in carbon-based polyacene~\cite{Cirera2020, jankowski2024excitonic}, graphene nanoribbons~\cite{Tepliakov2023}, and corresponds to the so-called obstructed topology, which in the spirit of the irrep considerations above, i.e., in terms of crystalline $\mathcal{I}$-symmetry indicators, is topologically captured by the fractional gluing conditions on the elementary band representations~\cite{kruthoff2017, Clas4, Clas5}.

\newpage
\section{Hyperpolarizability in terms of quantum metric}\label{app::B}

In the following, we derive the connection of the hyperpolarizability to quantum geometry. The derivation follows the physical steps of the derivation provided in Ref.~\cite{boyd}, with the key geometric insights introduced with the interpretations of transition amplitude products in terms of the elements of quantum geometric tensors~\cite{Ahn2021}.

We begin by noting that, as follows from the third-order perturbation theory in electron-photon coupling, $\Delta H = e \vec{\hat{r}} \cdot \vec{E}(t)$, written here in the length gauge, with $\vec{E}(t) = \vec{E}(\om) \text{cos}(\om t)$, the third-order susceptibility in response to optical fields $E_x(\om_i)$ compactly reads~\cite{boyd}:
\beq{eq:chi3dip}
    \chi^{(3)}_{xxxx}(\om; \om_1, \om_2, \om_3) = \frac{e^4}{\epsilon_0 a} \sum_{n,m,p,q,\kv} \mathcal{P}_F~\frac{r^x_{nm} r^x_{mp} r^x_{pq} r^x_{qn}}{(\hbar \om - E_{mn})(\hbar \om - E_{pn} - \hbar \om_2 - \hbar \om_3)(E_{qn} - \hbar \om_3) }.
\eeq
In the above, $\mathcal{P}_F$ denotes the permutation operator over all input and output frequencies~\cite{boyd}, $\epsilon_0$ is vacuum electric permittivity constant, and $r^x_{nm} = \bra{\psi_{n\kv}} \hat{x} \ket{\psi_{m\kv}}$ is the transition dipole matrix element per electron charge~\cite{Sipe2000, Ahn2021}. Here, we consider a coupling on combining the frequencies $\om_i$ with $i = 1,2,3$, which result in a response with frequency, $\om = \om_1 + \om_2 + \om_3$. The result in Eq.~\eqref{eq:chi3dip} is invariant from the choice of the length gauge. In velocity gauge the result would also be invariant, but the connection to the quantum geometry would become less transparent with an addition of associated energy factors. Therefore, the velocity gauge formulae are not considered further. The other contributions arising from the higher derivatives of the Bloch Hamiltonian,
that naturally occur in higher-order responses, are vanishing in the static limit $\om \rightarrow 0$, as elucidated in Ref.~\cite{boyd}, and consistently with the diagrammatic calculations of Ref.~\cite{Pronin1995}.

In terms of non-Abelian Berry connnection $A^x_{nm}$ matrix, we retrieve:
\beq{}
    \chi^{(3)}_{xxxx}(\om; \om_1, \om_2, \om_3) = \frac{e^4}{\epsilon_0 a} \sum_{n,m,p,q,\kv} \mathcal{P}_F~\frac{A^x_{nm} A^x_{mp} A^x_{pq} A^x_{qn}}{(\hbar \om - E_{mn})(\hbar \om - E_{pn} - \hbar \om_2 - \hbar \om_3)(E_{qn} - \hbar \om_3) }.
\eeq
As an example, the response associated with the Kerr nonlinearity~\cite{boyd} can be captured as,
\beq{}
    \chi_{xxxx}^{(3)}(\om; -\om, \om, \om) = \frac{e^4}{\epsilon_0 a} \sum_{n,m,p,q, \kv} \mathcal{P}_F~\frac{A^x_{nm} A^x_{mp} A^x_{pq} A^x_{qn}}{(\hbar \om - E_{mn})(- E_{pn} - \hbar \om)(E_{qn} - \hbar \om) }.
\eeq
On the other hand, as central to this work, the static hyperpolarizability $\chi^{(3)}$ in a one-dimensional system reads,
\beq{}
    \chi^{(3)} \equiv \chi_{xxxx}^{(3)}(\om \rightarrow 0; 0,0,0) = \frac{e^4}{\epsilon_0 a} \sum_{n,m,p,q, \kv} \frac{A^x_{nm} A^x_{mp} A^x_{pq} A^x_{qn}}{E_{mn} E_{pn} E_{qn}}.
\eeq
Hence, in terms of the non-Abelian quantum geometric tensor (QGT) defined in the main text, we arrive at: 
\beq{}
    \chi^{(3)} = \frac{e^4}{\epsilon_0 a} \sum_{n,m,p,q, \kv} \frac{[\textbf{Q}^{(p)}_{xx}]_{np} [\textbf{Q}^{(q)}_{xx}]_{pn} }{E_{mn}E_{pn}E_{qn}},
\eeq
where $\sum_\kv \equiv \frac{a}{2\pi} \int_\text{BZ} \dd k$ in one spatial dimension ($x$). Moreover, the considered non-Abelian QGT coincides with the non-Abelian quantum metric: ${\textbf{g}^{(p)}_{xx} = \text{Re}~ \textbf{Q}^{(p)}_{xx} = \textbf{Q}^{(p)}_{xx}}$. This culminates in the main result of the main text,
\beq{}
    \chi^{(3)} = \frac{e^4}{\epsilon_0 a} \sum_{n,m,p,q, \kv} \frac{[\textbf{g}^{(p)}_{xx}]_{np} [\textbf{g}^{(q)}_{xx}]_{pn} }{E_{mn}E_{pn}E_{qn}} = \frac{e^4}{\epsilon_0 } \int_\text{BZ} \frac{\dd k~}{2\pi} \sum_{n,m,p,q} \frac{[\textbf{g}^{(p)}_{xx}]_{np} [\textbf{g}^{(q)}_{xx}]_{pn} }{E_{mn}E_{pn}E_{qn}},
\eeq
which in the effective model central to the main text, we limited to the low-energy subspace $n,m,p,q=1,2$. We note that the dominant contribution to $\chi^{(3)}$ is expected to emerge from the crystalline topological bands $n=1, m=2$ which can arise by a band inversion and are neighboring the band gap. Correspondingly, we have: $E_{21} = \text{min}_{n \neq m}~|E_{mn}|$, which demonstrates that the low energy bands contributions dominate the band sum with the suppression of the denominator as: $\chi^{(3)} \sim E^{-3}_{21}$. Moreover, in the band-resolved picture, see Sec.~I, it is only the topological bands that are expected to  realize topologically enhanced quantum metric, providing for an additional enhancement $\chi^{(3)} \sim (g^{12}_{xx})^2$. The arguments above show that in the case of nontrivial crystalline band topology, it is precisely the bottom conduction and top valence bands that dominate amongst the other contributions to $\chi^{(3)}$.

\newpage
\section{Control of quantum geometry with external fields}\label{app::C}
We here consider the effects of the external fields, such as static  strain fields $\lambda$, to further enhance the quantum geometry and nonlinear optical properties for potential material applications. In the presence of external fields, the Bloch states can be written as $\ket{u_{n\kv}(\lambda)}$, and the multiband metric admits the external parameters as $g^{nm}_{xx}(\lambda)$.

While being close to the lower bound saturation is not desirable for optical material design, the following question of: how close to saturating the bound one can get by perturbing the system, nevertheless, arises. It is further interesting to consider the opposite limit of how away from the bound one can arrive, i.e., how big the metric can be. The metric should diverge in the gapless (metallic) limit, therefore small-gap materials with nontrivial quantum geometry are highly desirable for realizing the nonlinear optical responses. In terms of the velocity operator, $\hat{v}_x = \frac{i}{\hbar}[\hat{H}, \hat{x}]$, the quantum metric of interest reads,
\beq{}
    g^{12}_{xx}(\la) = \hbar^2 \frac{|\bra{\psi_{1\kv}(\la)} \hat{v}_x \ket{\psi_{2\kv}(\la)}|^2}{E^2_{21}(\la)},
\eeq
with $\ket{\psi_{n\kv}(\lambda)} = e^{i \kv \cdot \vec{r}} \ket{u_{n\kv}(\lambda)}$, the external parameter-dependent Bloch states. In terms of the scaling with the band energy difference $E_{21}(\la) = E_{2\kv}(\la) - E_{1\kv}(\la)$, with $E_{1\kv}(\la), E_{2\kv}(\la)$ the single-band energies, this relation reflects the local enhancement of the momentum space metric close to the direct band gap at the band edge ($k = \pi/a$). We retrieved such scaling in the main text, as shown in Fig.~2. The corresponding scaling is intuitively justified by the change of the overlaps of the exponentially localized basis (Wannier) orbitals. From the scaling of hoppings in the limit of high tensile strains ($\gamma_i \rightarrow \infty$), one obtains a large gap trivial atomic insulator ($E_g \rightarrow \infty$), topologically equivalent to vacuum. With the provided physical limit in mind, the practical starting point is to begin with a $\mathbb{Z}_2$-topological system, as the trivializations occur in the limit of large tensile strains. The tensile strains can trivialize the topology, while preserving the insulating character of the system $(E_g \neq 0)$, after the topological phase transition across a metallic state occurs.

We now provide more details on the controllable geometry based on the example realized in the strained Su-Schrieffer-Heeger (SSH) model~\cite{ssh1, ssh2} central to the main text. The introduced two-band model can be compactly rewritten in terms of $\mathfrak{su}(2)$ Lie algebra elements, the Pauli matrices, $\vec{\sigma}=(\sigma_x, \sigma_y,\sigma_z)$:
\begin{equation}
H(k,\lambda) = \vec{d}(k,\lambda) \cdot \boldsymbol{\sigma},    
\end{equation}
with the $\vec{d}$-vector:
\beq{}
    \vec{d}(k,\la) = 
    \begin{pmatrix}
    t_1(\la) + t_2(\la) \cos k a(\la) \\
    t_2(\la) \sin k a(\la) \\ 
    0
    \end{pmatrix}.
\eeq
In the real material context, the hopping parameters $t_1$ and $t_2$ can be retrieved by fitting the effective band structure of the model to the first principles calculations, i.e., to the DFT band structure. The electronic wave functions of valence/conduction electrons read:
\beq{}
    \ket{u_{1/2\kv}} = \frac{1}{\sqrt{2|\vec{d}(k,\la)|^2}} \begin{pmatrix} 
     {d_x(k, \la)} \pm \ii {d_y(k, \la)} \\ 
     |\vec{d} (k, \la)|
     \end{pmatrix},
\eeq
and the electronic single-particle energies are given by:
\beq{}
    E_{1/2 \kv}(\la) = \pm \sqrt{d^2_x(k,\la) + d^2_y (k, \la)} = \pm |\vec{d} (k, \la)|.
\eeq
Hence, the band energy differences read: $E_{21}(\la) = 2|\vec{d} (k, \la)|$.  The normalized vector: $\hat{\vec{d}}(k, \la) \equiv \vec{d}(k, \la)/|\vec{d} (k, \la)|$, with ${\hat{\vec{d}}(k, \la)\cdot  \hat{\vec{d}}(k, \la) =1}$, fully determines the quantum metric as
\beq{}
    g^{12}_{xx}(\la) = \frac{1}{4} \partial_{k} \hat{\vec{d}}(k, \la) \cdot   \partial_{k} \hat{\vec{d}}(k, \la).
\eeq
As a consequence, we can compactly rewrite the third-order susceptibility in the static limit, which corresponds to an external-parameter controllable hyperpolarizability:
\beq{} 
    \chi^{(3)}(\lambda) = \frac{e^4}{2\pi \epsilon_0} \int_{\text{BZ}}~\dd k~ \frac{|\partial_{k} \hat{\vec{d}}(k, \la)|^4}{128 |\vec{d} (k, \la)|^3}.
\eeq
The form compactly captures the geometric character of the hyperpolarizability, and its explicit dependence on strain $\la$, as captured by the vector $\hat{\vec{d}}(k, \la)$. The $\hat{\vec{d}}(k, \la)$ vector determines a measure of quantum distance with its tangent vector $\partial_{k} \hat{\vec{d}}(k, \la)$ over a great circle over a two-sphere $S^2$ in the parameter space of components $(d_x, d_y, d_z)$. This observation provides an intuitive connection between the hyperpolarizability and the notion of quantum distance definitional for the Fubini-Study metric~\cite{provost1980riemannian}: $ds^2 = g_{xx} \dd k \dd k$ over momentum space. 

It should be noted that the presence of additional bands can only increase, rather than reduce the left-hand side, as for any band indices $n,m$: $g^{nm}_{xx} \geq 0$, by definition, and in the multiband case: $\langle g_{xx} \rangle \equiv  \frac{a}{2\pi} \int_\text{BZ} \dd k~ \sum_{n,m} g^{nm}_{xx}$, with $n$ occupied and $m$ unoccupied bands. Hence, the quantum geometric bound in the main text holds for any arbitrary number of valence and conduction bands of gapped one-dimensional chains respecting the sublattice and inversion symmetries. Notably, if the chain is gapless: $|\vec{d}|=0$, and $\hat{\vec{d}} = \vec{d}/|\vec{d}|$ becomes ill-defined, while the hyperpolarizability in the metallic limit formally diverges.

\newpage
\section{Topological bound on hyperpolarizability}

We here derive a topological bound on the hyperpolarizability due to the (crystalline) band topology. We note that the variance of quantum metric over momentum space must be nonnegative,
\beq{}
    \text{Var}~\langle g_{xx} \rangle = \langle g^2_{xx} \rangle - \langle g_{xx} \rangle^2 \geq 0.
\eeq
Therefore, $\chi^{(3)} \sim \langle g^2_{xx} \rangle \geq \langle g_{xx} \rangle^2 \sim \nu_{\text{FK}}^2$, cf. the previous Section on hyperpolarizability. As a consequence, in the presence of the band gap, $\Delta = \text{min}_{\kv} [E_{m\kv} -E_{n\kv}]$, with $n$ running over all occupied and $m$ over unoccupied band indices, $\chi^{(3)}$ is topologically bounded as:
\beq{}
    \chi^{(3)} \geq \frac{e^4}{\epsilon_0} \frac{\langle g^2_{xx} \rangle}{\Delta^3} \geq \frac{e^4}{\epsilon_0} \frac{\langle g_{xx} \rangle^2}{\Delta^3} \geq \frac{e^4}{\epsilon_0} \frac{a^4}{16 \Delta^3} \nu^2_\text{FK}.
\eeq
Using the $f$-sum rule~\cite{Souza2000}, with $n_e \sim 1/a$, the ground state electron density given by the band filling,  and $m_e$ the electron mass, the gap is topologically bounded as~\cite{Kivelson1982, Onishi2024}:
\beq{}
    \frac{\hbar}{\Delta} \frac{n_e  e^2}{ 8 m_e a} = \frac{\hbar}{\Delta} \int^\infty_0 \dd \om~\sigma_{xx}(\om) \geq W_{xx} = \frac{\pi e^2}{\hbar} \frac{ \langle g_{xx} \rangle}{a} \geq \frac{\pi e^2 a}{4 \hbar} \nu_{\text{FK}},
\eeq
with $W_{xx}$ an optical weight defined in the main text. Hence, on rearranging, one obtains:
\beq{}
    \frac{1}{\Delta} \geq \frac{2\pi  m_e a^2}{n_e \hbar^2} \nu_{\text{FK}},
\eeq
which combined with the previous inequalities, culminates in:
\beq{}
    \chi^{(3)} \geq  \frac{\pi^3 e^4 m_e^3 a^{10}}{2 \epsilon_0 n_e^3 \hbar^6} \nu^{5}_\text{FK} = 6 \times 10^{50}~\text{m}^{-5}/\text{V}^2 \times a^{10}/n_e^3 .
\eeq
This fundamental condition shows that for nontrivial band topology, i.e.,~$\nu_\text{FK} = 1$, the hyperpolarizability is bounded from below by the fundamental constants, charge density, and by the unit cell lattice parameter $a$. In the cases of the other known topological gap bounds, such as by the Chern numbers~\cite{Onishi2024}, our argument directly generalizes to the other band topologies. Namely, on replacing the topological invariant, one obtains an analogous bound for any other invariant that realizes a gap bound in any other dimensionality of interest.

\newpage
\section{Multiband extension of the effective models}

We here detail the characteristics and responses of the effective Hamiltonians beyond the two-band low-energy theory, which mirror the band structures of realistic materials. We nonetheless show that while the additional band contributions are necessary for a quantitative match of the nonlinear responses with experiments~\cite{Bredas1994}, given such extended picture includes all possible virtual processes, the two-band models provide the dominant contributions to the considered hyperpolarizability. In Fig.~\ref{FigS1}, we show all results associated with the multiband models detailed in the next subsections.

\subsection{Polyacetylene}

We discuss an extended polyacetylene model, where more electronic bands can be included. We begin with the polyacetylene parameterization: $t_1 = 2.15~\text{eV}$, $t_2 = 2.85~\text{eV}$, $a_1 = 0.642~\AA$, $a_2 = 0.558~\AA$~\cite{ssh1, ssh2}, and set the decay parameters $\gamma_1 = 15 \AA^{-1}$, $\gamma_2 = 22 \AA^{-1}$ for the single and double bonds respectively. We consider a multiband extension of the polyacetylene model, where two more orbitals are involved (one per each sublattice), resulting in a $4 \times 4$ Hamiltonian:
\beq{}
    H(k) = 
    \begin{pmatrix}
    0&  0  & 0 &  t_3 + t_4 e^{ika}\\ 
     0& 0 & t_1 + t_2 e^{ika} & 0 \\
    0& t_1 + t_2 e^{-ika} & 0& 0\\
    t_3 + t_4 e^{-ika} &0 & 0& 0 \\
    \end{pmatrix},
\eeq
where $t_3 << t_1, t_2 << t_4$ are the additional couplings corresponding to the hoppings to higher energy orbitals. As long as $t_3 << t_1, t_2 << t_4$, which is justified, given the energetic separation of the $p_z$-constituted $\pi$~bands from the other $\sigma$~bands constituted by the $sp^2$ orbitals, the additional bands are less dispersive, with energies taking values far from Fermi level $\mu$, i.e., $E_{21} << E_{31}, E_{42}$, which provides for a cubic suppression of the corresponding hyperpolarizability contribution, with negligible quantum metric contributions, which also scale with the band gap, cf.~Sec.~II of the Supplemental Material. In Fig.~\ref{FigS1}, we set $t_3 = -0.4~\text{eV}$, $t_4 = 5~\text{eV}$. We arrive at the same conclusions for topological bounds.

\subsection{Poly(p-phenylene)}

We detail a multiband model of poly(p-phenylene) (PPP), as an intermediate case between polyacetylene and polypentacene addressed in the next subsection. In the multiband extension, we assume that each phenylene ring contributes three occupied molecular orbitals (MOs) of benzene, which results in an effective $6 \times 6$ model Hamiltonian,
\beq{}
    H(k) = 
    \begin{pmatrix}
    \epsilon_2 & t_5 + t_6 e^{ika} & 0 & 0 & 0 & 0\\
   t_5 + t_6 e^{-ika} & \epsilon_1 &   0 &  0  & 0 & 0 \\ 
    0&  0& 0 & t_1 + t_2 e^{ika} & 0 &0\\
   0 & 0& t_1 + t_2 e^{-ika} & 0& 0 & 0\\
   0& 0 &0 & 0& \epsilon_2 & t_3 + t_4 e^{ika}  \\
  0 & 0 & 0 & 0 & t_3 + t_4 e^{-ika} & \epsilon_1\\
    \end{pmatrix},
\eeq
with $t_4, t_6 << t_1, t_2 << t_3, t_5$, where $t_5, t_6$ are the hoppings to/from the lower energy benzene molecular orbitals (MOs), with energies ($\epsilon_1, \epsilon_2 < 0$). We find that the HOMOs of benzene molecules, hoppings between which define $t_1, t_2$, contribute the most of the quantum metric. The lower energy orbitals that define $t_3, t_4, t_5, t_6$ provide for additional bands with larger energy differences, and hence experience higher energy gaps than $E_{21}$. The higher energy gaps suppress the hyperpolarizability contributions due to the additional bands in a cubic manner. For the purposes of Fig.~\ref{FigS1}, we set $\epsilon_1 = -2.5~\text{eV}$, $\epsilon_2 = -1.5 ~\text{eV}$, $t_1 = -0.65~\text{eV}$, $t_2 = -3.0~\text{eV}$, $t_3 = 0.1~\text{eV}$, $t_4 = 0.0~\text{eV}$, $t_5 = -1.0~\text{eV}$, $t_6 = 0.7~\text{eV}$. 

\subsection{Polypentacene}

We detail a polyacene model, in particular for polypentacene, which is known to be topologically ($\mathbb{Z}_2$) nontrivial~\cite{Cirera2020}, and which was previously modelled with a two-band effective theory, similarly to as we adapt in the main text. For a multiorbital extension of the model, one can define a $6 \times 6$ Hamiltonian, where electrons can hop between molecular orbitals contributed by linear combinations of five rings $(t_1, t_3, t_5)$, i.e., between different orbitals within a monomer unit, as well as from pentacene rings across the acetylene unit $t_2, t_4, t_6$. The modified tight-binding Hamiltonian including internal hoppings between different orbitals of pentacene, reads:
\beq{}
    H(k) = 
    \begin{pmatrix}
    \epsilon_2 & 0 & 0 & 0 & 0 & t_5 + t_6 e^{ika}\\
  0 & \epsilon_1 &   0 &  0  & t_3 + t_4 e^{ika} & 0 \\ 
    0&  0& 0 & t_1 + t_2 e^{ika} & 0 &0\\
   0 & 0& t_1 + t_2 e^{-ika} & 0& 0 & 0\\
   0& t_3 + t_4 e^{-ika} &0 & 0& \epsilon_2 & 0  \\
   t_5 + t_6 e^{-ika} & 0 & 0 & 0 & 0 & \epsilon_1\\
    \end{pmatrix},
\eeq
with onsite energies of the lower-energy pentacene MOs employed as the tight-binding basis functions, $\epsilon_1, \epsilon_2 < 0$, and hoppings $t_2 > t_1$ in the topological $\mathbb{Z}_2 = 1$ phase.

We find that even in the presence of additional electronic bands, the hyperpolarizability of the material is predominantly determined by the low-energy two-band subspace. This is to be expected, as the effective hoppings induce a significant dispersion only in the topological valence/conduction bands. As a result, the additional bands are less dispersive, realize marginal additional quantum geometric contributions, and are separated by larger gaps. Hence, correspondingly, both the numerator and denominator of the hyperpolarizability contributions due to these bands, are negligibly small, as long as $\mathbb{Z}_2 = 1$, which is consistent with the effective polypentacene bands retrieved in literature~\cite{Cirera2020, jankowski2024excitonic}. For the purposes of Fig.~\ref{FigS1}, we set $\epsilon_1 = -2.5~\text{eV}$, $\epsilon_2 = -2.5 ~\text{eV}$, $t_1 = 0.3~\text{eV}$, $t_2 = 0.5~\text{eV}$, $t_3 = 0.5~\text{eV}$, $t_4 = 1~\text{eV}$, $t_5 = 0.1~\text{eV}$, $t_6 = 0.9~\text{eV}$.

\subsection{Graphene nanoribbons}

Further to the polyacetylene~\cite{ssh1, ssh2}, PPP, and polypentacene~\cite{Cirera2020}, we consider the graphene nanoribbons (GNRs), which are known to host a strain-dependent crystalline topology protected by inversion symmetry~\cite{Tepliakov2023}. To minimally capture the phenomenology of GNRs, one can consider a set of coupled $4 \times 4$ Bloch Hamiltonians due to $n$ (\textit{cis}) $\pi$-bonded chains with a four-orbital basis. The corresponding models enjoy a $U(2n)$ gauge freedom in the occupied band manifold. In such case, the parameter space for realizing nontrivial Bloch state geometry is further spanned by a set of strain parameters $\gamma_i$, with $\gamma_i = 1,2,3$  separately tunable for hoppings $t_i = t_1, t_2, t_3$.

\begin{figure}
\includegraphics[width=\columnwidth]{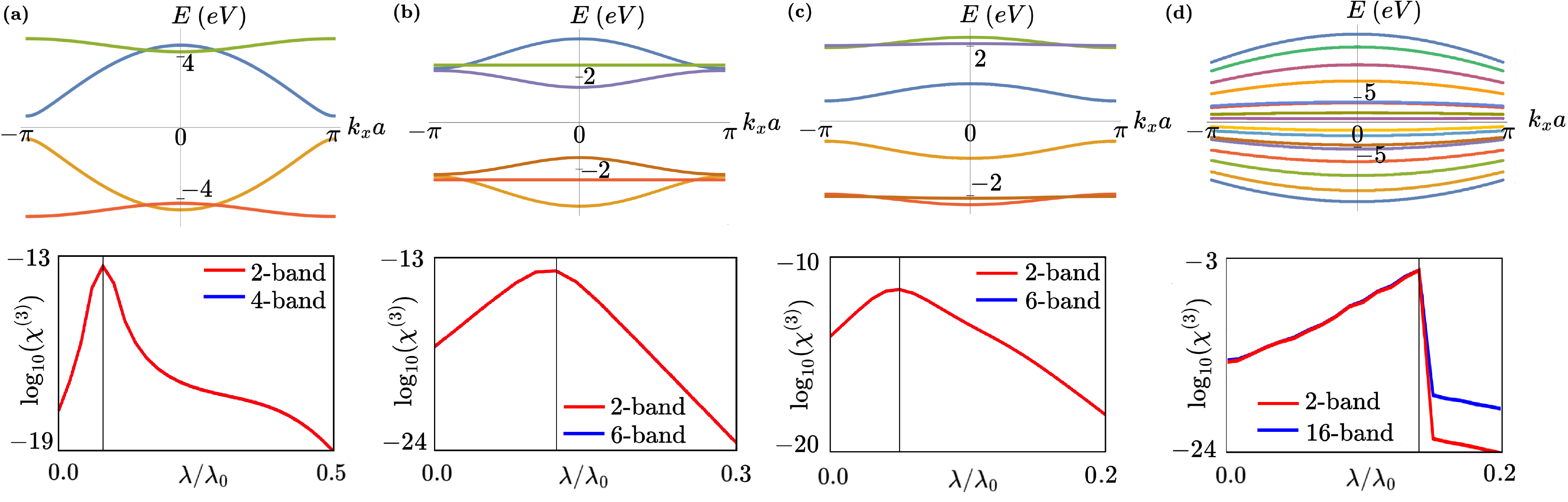}
\caption{
{The band structures and hyperpolarizabilities of the multiband extensions of $\pi$-conjugated polymer models.} {(a)}~polyacetylene [two $\pi$ bands and two $\sigma$ bands],  {(b)} poly(p-phenylene) [six $\pi$ bands],  {(c)} polypentacene [six $\pi$ bands], {(d)} graphene nanoribbon [sixteen $\pi$~bands]. The tight-binding parameters were approximately fitted to the band structures of materials with parametrizations provided in the corresponding subsections. We overlay the approximate two-band model hyperpolarizabilities over multiband hyperpolarizabilities $\chi^{(3)}$ obtained within the extended models. As anticipated with the intuitive $\chi^{(3)}$ scaling argument, we observe that in the topological phases, the hyperpolarizabilities change marginally on the logarithmic scale after having included the additional lower energy bands.}
\label{FigS1}
\end{figure}

The scaling of the hoppings and tensile strains $\varepsilon_i > 0$, with $\varepsilon_i \equiv (a_i(\lambda)/a_i-1)$~\cite{Tepliakov2023}, and $a_i(\lambda)$ strain-dependent lattice hopping distances, can be minimally captured within the considered models by a substitution: $t_i \rightarrow t_i e^{- \gamma_i \varepsilon_i }$, where $\gamma_i$ reflect bond-dependent decay rates, which depend on the strains and on the geometry of the orbitals between which the hopping occurs. The hoppings $t_1$, $t_2$, $t_3$ correspondingly occur over nearest-neighbor (NN), next-nearest-neighbor (NNN), and next-next-nearest-neighbor (NNNN) vectors with the $x$-direction-projected lengths $|\vec{a}_1|=a_0/2$, $|\vec{a}_2|=\sqrt{3}a_0/2$, $|\vec{a}_3| = a_0$, with $a_0 = 3 \AA$. The $4 \times 4$ chain Hamiltonian subblock reads,
\begin{small}
\beq{}
    H^{(4)}(k, \Delta_1, \Delta_2) = 
    \begin{pmatrix}
    0& 2t_1 \cos ka_1 & 2t_2 \cos ka_2 & t_1 e^{ika_1} + (t_1 + \Delta_2) e^{-ika_1}   + t_3 e^{ika_3} \\ 
    2t_1 \cos ka_1& 0& t_1 e^{ika_1} + (t_1+\Delta_1) e^{-ika_1} &2t_2 \cos ka_2 \\
    2t_2 \cos ka_2& t_1 e^{-ika_1} + (t_1+\Delta_1) e^{ika_1} & 0& 2 t_1 \cos k a_1 \\
    t_1 e^{-ika_1} + (t_1 + \Delta_2) e^{ika_1}   + t_3 e^{-ika_3} &2t_2 \cos ka_2 & 2 t_1 \cos k a_1& 0
    \end{pmatrix},
\eeq
\end{small}
where $\Delta_1, \Delta_2$ are the hopping imbalances at both GNR edges, respectively. The hopping imbalance is consistent with the picture of Ref.~\cite{Tepliakov2023}. Furthermore, the $4 \times 4$ chain subblocks are additionally coupled by $4 \times 4$ interchain coupling matrices:
\beq{}
    H^{(c)}(k) = 
    \begin{pmatrix}
    0& 0 & 0 & 0\\ 
    2t_1 \cos ka_1& 0&0 & 2t_2 \cos ka_2 \\
    2t_2 \cos ka_2& 0 & 0&  2t_1 \cos ka_1\\
    0 &0 & 0& 0 \\
    \end{pmatrix}.
\eeq
On combining the $4 \times 4$ chain Hamiltonian blocks with the interchain coupling blocks, we arrive at:
\beq{}
    H^{(16)}(k, \Delta) = 
    \begin{pmatrix}
    H^{(4)}(k, \Delta_1 = \Delta, \Delta_2=0)& H^{(c)}(k) & 0 & 0\\ 
    H^{(c)}(k)& H^{(4)}(k, \Delta_1=0, \Delta_2=0)&H^{(c)}(k) &0 \\
    0& H^{(c)}(k)& H^{(4)}(k, \Delta_1=0, \Delta_2=0)& H^{(c)}(k)\\
    0 &0 & H^{(c)}(k)& H^{(4)}(k, \Delta_1=0, \Delta_2 = \Delta) \\
    \end{pmatrix},
\eeq
which is definitionally consistent with the models of Ref.~\cite{Tepliakov2023}, given the identical real space hoppings. For the purposes of Fig.~\ref{FigS1}, we set $\Delta = -0.2~\text{eV}$, $t_1 =-2.88 ~\text{eV}$, $t_2 = 0.22~\text{eV}$, $t_3 = -0.25 ~\text{eV}$~\cite{Tepliakov2023}.

\newpage

\section{Anharmonic classical oscillator in the inversion-symmetric potential}\label{app::E}
In the following, we provide further details on the physics of an anharmonic classical oscillator in the inversion-symmetric potential within a semiclassical relation to the quantum geometry central to the main text. We begin with an anharmonic classical Hamiltonian with the lowest-order inversion-symmetric anharmonicity introduced in the main text: $H = \frac{p^2}{2m} + U(x) = \frac{p^2}{2m} + \frac{1}{2} kx^2 + \frac{1}{4} b x^4$, obtaining:
\beq{}
    \dot{p} = -kx - bx^3,
\eeq
from the Hamilton equation of motion $\dot{p} = -\partial H /\partial x$. The other equation of motion obtains: $\dot{x} =\partial H /\partial p = p/m$. The spring constant can be moreover written as $k = m \om_0^2$, with $\om_0$ the natural frequency of the oscillator. 

We now focus on the semiclassical picture associated with the anharmonic oscillator -- from the unique perspective of quantum geometry central to this work. The polarization arises from the displacement of an oscillating charge
\beq{}
    P(t) = -e~\delta x(t),
\eeq
where $\delta x(t)$ is a time-dependent charge displacement under the adiabatic forcing $F$. The forcing is equivalent to a perturbation:
\beq{}
    \Delta H = Fx\cos \om t = -eE_0 x\cos \om t,
\eeq 
which, on changing the Hamiltonian accordingly: $H \rightarrow H + \Delta H$, obtains the following equation of motion:
\beq{}
    m \ddot{x} + kx + bx^3 = -eE_0~ \text{cos}~\om t.
\eeq
To proceed, we consider a perturbative solution of form $x(t) = \la_p x^{(1)}(t) + \la_p^2 x^{(2)}(t) + \la_p^3 x^{(3)}(t)$, with $x^{(n)}$ representing an oscillator component with frequency $n \om$ and $\la_p$ representing a parameter of the perturbative expansion. We note that $x^{(2)}(t) = 0$ vanishes by inversion symmetry. On investigating the first and third order in $\la_p$, we have:
\beq{}
    \ddot{x}^{(1)} + \om^2_0 x^{(1)} = -\frac{eE_0}{m}~ \text{cos}~\om t,
\eeq
and at the third perturbative order,
\beq{eq:third_diff}
    \ddot{x}^{(3)} + \om^2_0 x^{(3)} - \frac{b}{m} (x^{(1)})^3= -\frac{eE_0}{m}~ \text{cos}~\om t.
\eeq
We assume steady state solutions of form:
\beq{}
    x^{(1)}(t) = \int \dd \om~ x^{(1)}(\om) e^{-i\om t},
\eeq
\beq{}
    x^{(3)}(t) = \int \dd \om'~ x^{(3)}(\om') e^{-i\om' t},
\eeq
which upon insertion into the equations of motion obtains:
\beq{}
    x^{(1)}(\om) = \frac{-eE_0}{m(\om_0^2 - \om^2)} \equiv \frac{-eE_0(\om)}{m(\om_0^2 - \om^2)}.
\eeq
On inserting into the third-order displacement equation, we have:
\beq{}
    \ddot{x}^{(3)} + \om^2_0 x^{(3)} - \frac{b}{m} (x^{(1)})^3= \ddot{x}^{(3)} + \om^2_0 x^{(3)} -\int \int \int \dd \om~\dd \om'~\dd \om''~\frac{b e^3 E_0(\om) E_0 (\om') E_0 (\om'')}{m^4(\om_0^2 - \om^2)(\om_0^2 - \om'^2)(\om_0^2 - \om''^2)} e^{-i(\om+\om'+\om'')t},
\eeq
which on solving the differential equation [Eq.~\eqref{eq:third_diff}] culminates in:
\beq{}
    x^{(3)}(\om) = -\int \int \int ~\dd \om'~\dd \om''~\dd \om'''~\frac{b e^3  E_0 (\om') E_0 (\om'') E_0 (\om''')}{m^4 (\om_0^2 - \om^2)(\om_0^2 - \om'^2)(\om_0^2 - \om''^2)(\om^2_0 - \om'''^2)}.
\eeq
On comparing with the definition of the third-order susceptibility:
\beq{}
    P^{(3)}(\om) = -e~x^{(3)}(\om) = \epsilon_0 \int~\dd\om' \int~\dd\om'' \int~\dd\om'''~\chi^{(3)}_{xxxx}(\om, \om', \om'', \om''') E_0 (\om') E_0 (\om'') E_0 (\om'''),
\eeq
we arrive at the third-order susceptibility per electron,
\beq{}
   \chi^{(3)}_{xxxx}(\om, \om', \om'', \om''') = \frac{b e^4}{\epsilon_0 m^4(\om_0^2 - \om^2)(\om_0^2 - \om'^2)(\om_0^2 - \om''^2)(\om_0^2 - \om'''^2)}.
\eeq
On setting a static limit ($\om, \om', \om'', \om''' \rightarrow 0$) corresponding to the hyperpolarizability of interest:
\beq{}
    \chi^{(3)} = \frac{b e^4}{\epsilon_0 m^4 \om^8_0}.
\eeq
We now employ the interpretation of the averaged quantum metric as effective spring constant, $k = m \om_0^2 \sim 1/\langle g_{xx} \rangle$~\cite{Onishi2024}. We furthermore estimate: $b/m \approx \om_0^2 /\langle g_{xx} \rangle$, as effectively, the anharmonic and restoring forces become comparable at the distance~$d$: $b d^3 \sim kd = m\om^2_0 d$, with the lengthscale $d$ determined by the quantum metric weight $d^2 \sim \langle g_{xx} \rangle$, as given by the spread of the maximally localized Wannier functions~\cite{Marzari1997}. Hence, we have:
\beq{}
    \chi^{(3)} = \frac{e^4}{\epsilon_0} \frac{b}{m \om^2_0} \frac{1}{k^3} \sim  \frac{e^4}{\epsilon_0} \frac{1}{\om^2_0} \frac{b}{m} \frac{1}{k^3} \sim \frac{e^4}{\epsilon_0} \frac{1}{\om^2_0} \frac{\om^2_0}{\langle g_{xx} \rangle} \langle g_{xx} \rangle^3 \sim \frac{e^4}{\epsilon_0} {\langle g_{xx} \rangle}^2, 
\eeq
which concludes the derivation of the effective hyperpolarizability scaling with the averaged quantum metric [$\chi^{(3)} \sim {\langle g_{xx} \rangle}^2$] mentioned in the main text.

\end{widetext}

\clearpage

\bibliographystyle{apsrev4-1}
\bibliography{references.bib}